\documentclass{article}

\usepackage{arxiv}

\usepackage[T1]{fontenc}    
\usepackage{url}            
\usepackage{booktabs}       
\usepackage{amsfonts}       
\usepackage{nicefrac}       
\usepackage{microtype}      
\usepackage{lipsum}
\usepackage{graphicx}
\usepackage{subfigure}
\usepackage{tabularx}
\usepackage{array}
\usepackage{caption} 
\RequirePackage{amsmath,amsfonts,amssymb}
\RequirePackage{mathptmx}

\RequirePackage{xcolor}
\RequirePackage{authblk}
\RequirePackage[latin1,utf8]{inputenc}
\RequirePackage[english]{babel}
\RequirePackage{lmodern}
\RequirePackage{siunitx}
\RequirePackage{textgreek}

\RequirePackage{gensymb}
\RequirePackage{textcomp}
\RequirePackage[version=4]{mhchem}
\RequirePackage[scaled]{helvet}
\RequirePackage[T1]{fontenc}
\RequirePackage{lettrine} 
\RequirePackage[rightcaption]{sidecap} 
\RequirePackage[misc]{ifsym} 
\RequirePackage{bbding} 

\DeclareUnicodeCharacter{00B0}{\degree}

\RequirePackage[colorlinks=true, allcolors=blue]{hyperref}

\RequirePackage{cleveref}

\title{An ensemble-based approach by fine-tuning the deep transfer learning models to classify pneumonia from Chest X-ray images}

\author{
  Sagar Kora Venu \\
  Department of Analytics and Data Science\\
  Harrisburg University of Science and Technology\\
  Harrisburg, PA 17101 \\
  \texttt{SKora@my.harrisburgu.edu}
}

\usepackage{graphicx}
\usepackage[printonlyused,nohyperlinks]{acronym}
\usepackage{comment}
\newcommand\MyBox[1]{%
    \fbox{\parbox[c][3cm][c]{3cm}{\centering #1}}%
}
\newcommand\MyVBox[1]{%
    \parbox[c][1cm][c]{1cm}{\centering\bfseries #1}%
}  
\newcommand\MyHBox[2][\dimexpr3cm+2\fboxsep\relax]{%
    \parbox[c][1cm][c]{#1}{\centering\bfseries #2}%
}  
\newcommand\MyTBox[4]{%
    \MyVBox{#1}
    \MyBox{#2}\hspace*{-\fboxrule}%
    \MyBox{#3}\par\vspace{-\fboxrule}%
}  
\newcommand*\rot{\rotatebox{90}}

\begin{document}
\maketitle

\begin{abstract}
Pneumonia is caused by viruses, bacteria, or fungi that infect the lungs, which, if not diagnosed, can be fatal and lead to respiratory failure. 
More than 250,000 individuals in the United States, mainly adults, are diagnosed with pneumonia each year, and 50,000 die from the disease. 
Chest Radiography (X-ray) is widely used by radiologists to detect pneumonia. It is not uncommon to overlook pneumonia detection for a well-trained radiologist, which triggers the need for improvement in the diagnosis's accuracy. 
In this work, we propose using transfer learning, which can reduce the neural network's training time and minimize the generalization error. 
We trained, fine-tuned the state-of-the-art deep learning models such as InceptionResNet, MobileNetV2, Xception, DenseNet201, and ResNet152V2 to classify pneumonia accurately. Later, we created a weighted average ensemble of these models and achieved a test accuracy of 98.46\%, precision of 98.38\%, recall of 99.53\%, and f1 score of 98.96\%. 
These performance metrics of accuracy, precision, and f1 score are at their highest levels ever reported in the literature, which can be considered a benchmark for the accurate pneumonia classification.
\end{abstract}


\section{Introduction}

Pneumonia is an acute respiratory infection caused by bacteria, fungi, or viruses with mild to life-threatening conditions that, if not diagnosed, can lead to respiratory failure \cite{Pneumoni94:online}, \cite{Pneumoni23:online}. 
More than 250,000 individuals in the United States, mainly adults, are diagnosed with pneumonia each year, 50,000 die from the disease \cite{Pneumoni23:online}. 
Pneumonia is also the world's largest infectious cause of child mortality, accounting for 15\% of all infant deaths under five years of age \cite{Pneumoni94:online}. 
Standard tests for pneumonia diagnosis include blood tests, chest X-rays, pulse oximetry, sputum tests, arterial blood gas tests, bronchoscopy, pleural fluid culture, and CT scans \cite{Pneumoni73:online}. 
However, chest X-rays are a gold standard tool for diagnosing pneumonia that can distinguish pneumonia from other respiratory infections \cite{mandell2007infectious}. 
It is not uncommon to overlook pneumonia detection for a well-trained radiologist, which triggers the need for improvement in the diagnosis's accuracy.

Deep learning is now the state-of-the-art paradigm of machine learning, leading to enhanced performance in various areas, including medical image classification, natural language processing, object detection, segmentation, and other tasks \cite{litjens2017survey}, \cite{shen2017deep}, \cite{lundervold2019overview}. 
In particular, the deep Convolutional Neural Nets (CNN), which almost halve the error rate in the competition for Object Recognition - the Imagenet Large Scale Visual Recognition Competition (ILSVRC), have been highly dominant in field of computer vision \cite{krizhevsky2012imagenet}. 
Following CNN's success with computer vision, the medical image analysis community started to recognize the potential of deep learning techniques to achieve an expert level of performance in classification, segmentation, and detection of medical images \cite{litjens2017survey}. 
This work's significant contribution is that we propose the weighted average ensemble-based approach by fine-tuning the deep transfer learning models (InceptionResNet, MobileNetV2, Xception, DenseNet201, ResNet152V2) to improve the deep learning classification model's performance metrics.

The rest of the paper structure is as follows: In Section \ref{Related Work}, we briefly review the literature. Section \ref{Methods and Materials} introduces deep learning methods and the materials used and referred to throughout the study. 
Section \ref{Results} discusses the results obtained, and Section \ref{Conclusions} concludes the research.

\section{Related Work} \label{Related Work}

The deep learning framework proposed by Liang et al. \cite{liang2020transfer} incorporates transfer learning combined with residual thought and dilated convolution for the classification of pediatric pneumonia images, achieved a test recall of 96.7\%, and an f1 score of 92.7\%. 
To classify pneumonia from chest X-ray images, Chouhan et al. \cite{chouhan2020novel} and Hashmi et al. \cite{hashmi2020efficient} used transfer learning and proposed an ensemble model that combined the pre-trained models' results, achieving 96.4\% accuracy and 98.43\% accuracy respectively from the unseen dataset of the Guangzhou Women and Children's Medical Center. 
Stephen et al. \cite{stephen2019efficient} trained a convolutional neural network from scratch to detect the presence of pneumonia from a series of chest X-ray images resulting in approximately 94\% validation accuracy. 
Rahman et al. \cite{rahman2020transfer} used transfer learning from DenseNet201 architecture, a pre-trained deep convolutional network on the Imagenet dataset, and reported a 98\% accuracy of pneumonia classification. 
Ayan et al. \cite{ayan2019diagnosis} used Xception and Vgg16 as transfer learning models and compared the accuracy between them only to report the accuracy of the Xception network exceeds the Vgg16 network at 87\% and 82\%.

This paper's significant contribution is using a weighted average ensemble method by fine-tuning the state-of-the-art pre-trained neural networks trained on the Imagenet dataset to achieve the best classification performance metrics ever published in the literature.

\section{Methods and Materials} \label{Methods and Materials}
\begin{figure}[htbp]
\centering
\includegraphics[scale=0.4]{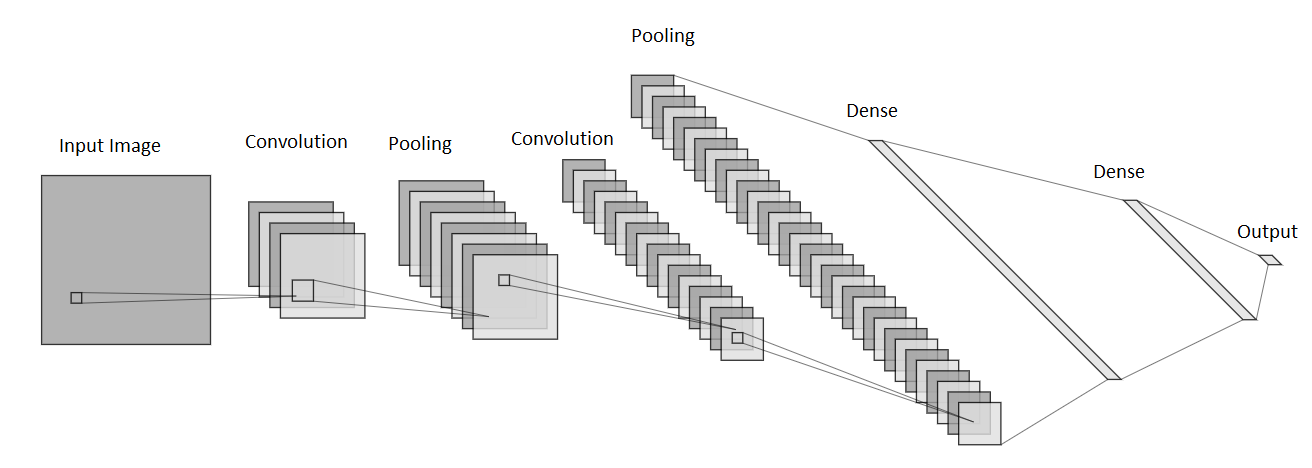}
\caption{CNN Architecture}
\label{fig:CNN_Architecture}
\end{figure}

Convolutional Neural Networks are a type of deep learning models designed for processing data in the form of multiple arrays, e.g., a color image has three channels (RGB), each channel consists of 2D arrays containing pixel intensities \cite{lecun2015deep}. 
The architecture of typical Convolutional Neural Network is shown in Figure~\ref{fig:CNN_Architecture}.

The first few stages in the architecture are a series of convolution layers and pooling layers. The image is fed as an input to the convolution layer to extract meaningful features (feature maps). 
A non-linearity is applied to the feature maps, followed by a pooling layer that merges similar features into one by computing either the maximum or average value for each patch on the feature map, which typically reduces the representation's dimensions. 
The output from the last stage of the convolution layer, non-linearity, and pooling layer is subjected to fully-connected layers, followed by a softmax to output the predictions.

\subsection{Transfer Learning} \label{Transfer Learning}
Machine learning algorithms assume that training and test data will come from the same distribution and feature space \cite{pan2009survey}. 
It may not hold good in real-world applications, particularly in the field of medical imaging, where obtaining a huge amount of training data is itself a major bottleneck due to high annotation costs and the protection of patients' privacy. 
Transfer Learning, which is a technique that improves the learning in a new domain through the transfer of knowledge from a related domain \cite{weiss2016survey}, \cite{torrey2010transfer}, bypasses the assumption that the training data must be independent and identically distributed (i.i.d) with the test data \cite{tan2018survey}.

\subsection{Pre-trained Image Classification Models} \label{Pre-trained Image Classification Models}
Pre-trained models are the models trained on large benchmark datasets, where the models have already learned  to extract a wide variety of features, can be used as a starting point to learn on a new task in a related domain. 
It is a common practice in the field of computer vision to use transfer learning via pre-trained models. 
In the following sub-sections, we will briefly introduce the pre-trained models used in this study.
\subsubsection{Xception} \label{Xception}
Xception is one of the state-of-the-art deep learning model architectures, based on depthwise separable convolution layers developed by Chollet \cite{chollet2017xception} from Google Inc, which is also known as the extreme version of Inception. 
The depthwise separable convolution consists of a depthwise convolution - a spacial convolution performed independently across every input channel, followed by a pointwise convolution - a 1 x 1 convolution that changes the input dimensions. 
But the extreme form of the inception module consists of a pointwise convolution followed by a depthwise convolution, and another difference among them is the presence/ absence of the non-linearity layer. 
Usually, depthwise separable convolutions are implemented without non-linearities between a depthwise convolution and pointwise convolution. 
In the extreme version of the inception module, depthwise convolution and pointwise convolution are followed by a ReLU non-linearity.

\begin{figure}[htbp]
\centering
\includegraphics[scale=0.5]{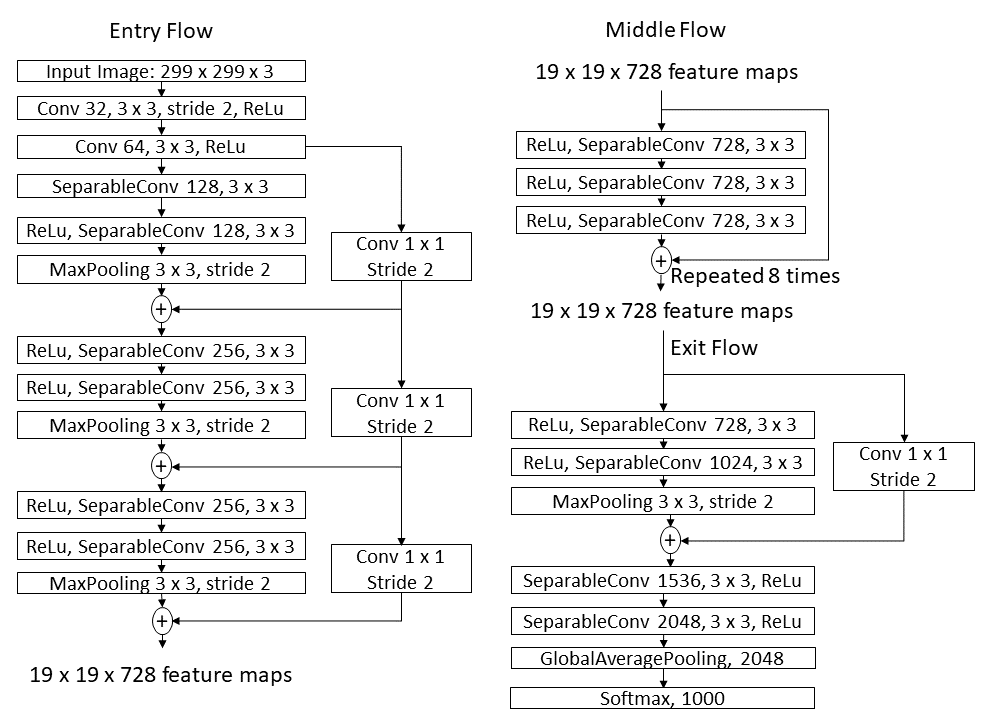}
\caption{Xception Architecture}
\label{fig:Xception_Architecture}
\end{figure}

The Xception architecture is shown in Figure~\ref{fig:Xception_Architecture}, which is divided into three major phases: Entry flow, Middle Flow, and Exit flow. There are 36 convolution layers in the architecture that are structured into 14 modules. 
Except for the first and last modules, all other modules have linear residual connections around them. 
In other words, Xception architecture is a linear stack of depthwise separable convolutions with residual connections, when trained on ImageNet dataset \cite{russakovsky2015imagenet}, Chollet \cite{chollet2017xception} reported a top-1 accuracy of 79.0\% and top-5 accuracy of 94.5\%. 

\subsubsection{MobileNetV2} \label{MobileNetV2}
\begin{figure}[htbp]
     \begin{center}
        \subfigure[MobileNet-V2 Architecture]{%
            \label{fig:MobileNet-V2-Architecture}
            \includegraphics[scale=0.5]{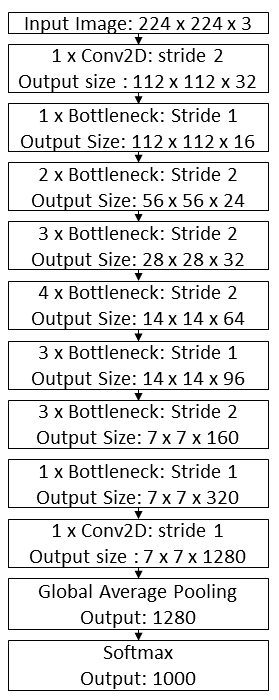}
        }%
        \quad\quad\quad\quad\quad\quad\quad
        \subfigure[MobileNet-V2 Bottleneck Stride Block]{%
           \label{fig:MobileNet-V2-Bottleneck-Stride-Block}
           \includegraphics[scale=0.4]{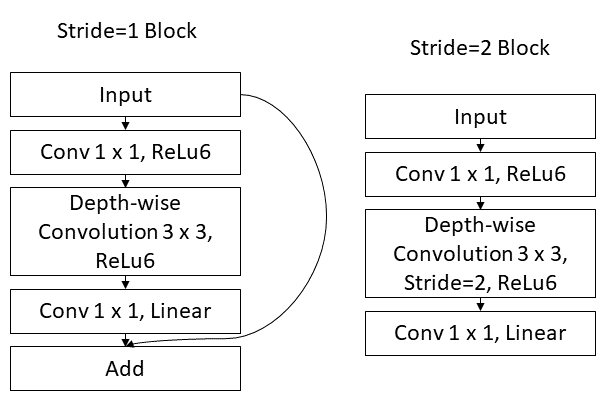}
        }\\ 
    \end{center}
    \caption{%
        MobileNet-V2 Architecture and Bottleneck Stride Blocks
     }%
   \label{fig:MobileNet-V2-Architecture-and-Bottleneck-Stride-Blocks}
\end{figure}

Sandler et al. \cite{sandler2018mobilenetv2} have introduced a neural network architecture designed specifically for mobile and resource-intensive environments. 
They introduced a unique layer module known as the inverted residual with a linear bottleneck, which takes a low dimensional compressed representation as an input that is then expanded to a high dimension and later filtered with a lightweight depth-wise convolution. 
The MobileNet-V2 architecture is shown in Figure~\ref{fig:MobileNet-V2-Architecture-and-Bottleneck-Stride-Blocks} that contains an initial fully convolutional layer followed by residual bottleneck layers. 

There are two types of blocks in the network, as shown in Figure~\ref{fig:MobileNet-V2-Bottleneck-Stride-Block}: one is the residual block of stride 1, and another is a block with stride 2 for downsizing the input from the previous layer. 
Each block has three layers: The first layer is a 1 x 1 Convolution with ReLu6 activation, the second layer is a depth-wise convolution, which is responsible for performing lightweight filtering by applying a single convolutional filter per input channel, and the third layer is a 1 x 1 Convolution, which is also referred to as a point-wise convolution that creates new features through computing linear combinations of the input channels. 
With this architecture, Sandler et al. \cite{sandler2018mobilenetv2} trained a neural network model on the ImageNet dataset \cite{russakovsky2015imagenet} and compared the performance with other similar mobile models: ShuffleNet and NasNet-A, and reported a top-1 accuracy of 74.7\% with ShuffleNet at 73.7\% and NasNet-A at 74.0\%.

\subsubsection{InceptionResNet} \label{InceptionResNet}

\begin{figure}[htbp]
     \begin{center}
        \subfigure[Large scale schema structure]{%
            \label{fig:Large-scale-schema-structure}
            \includegraphics[scale=0.5]{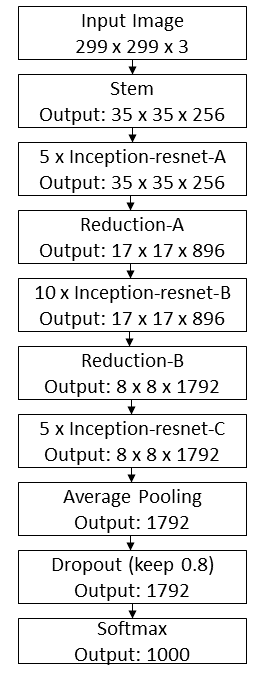}
        }%
        \quad\quad\quad\quad\quad\quad\quad\quad\quad\quad
        \subfigure[Schema for Stem module]{%
           \label{fig:Schema-for-Stem-module}
           \includegraphics[scale=0.5]{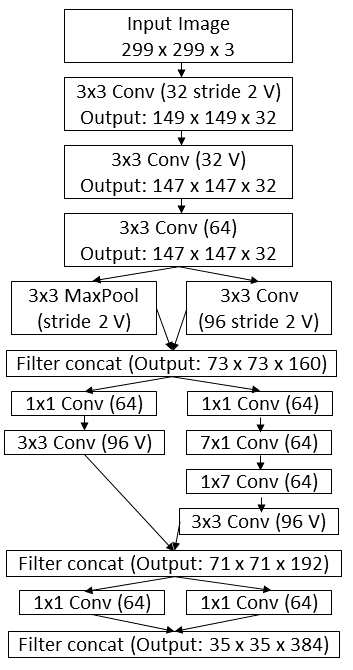}
        }\\ 
        \subfigure[Schema for Inception-ResNet modules]{%
            \label{fig:Schema-for-Inception-ResNet-modules}
            \includegraphics[scale=0.5]{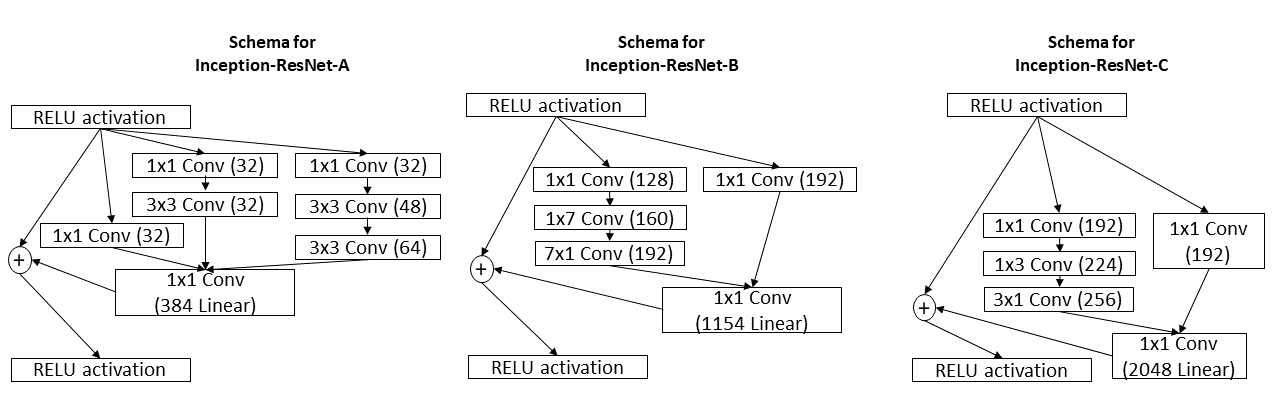}
        }\\ 
        \subfigure[Schema for Reduction modules]{%
            \label{fig:Schema-for-Reduction-modules}
            \includegraphics[scale=0.45]{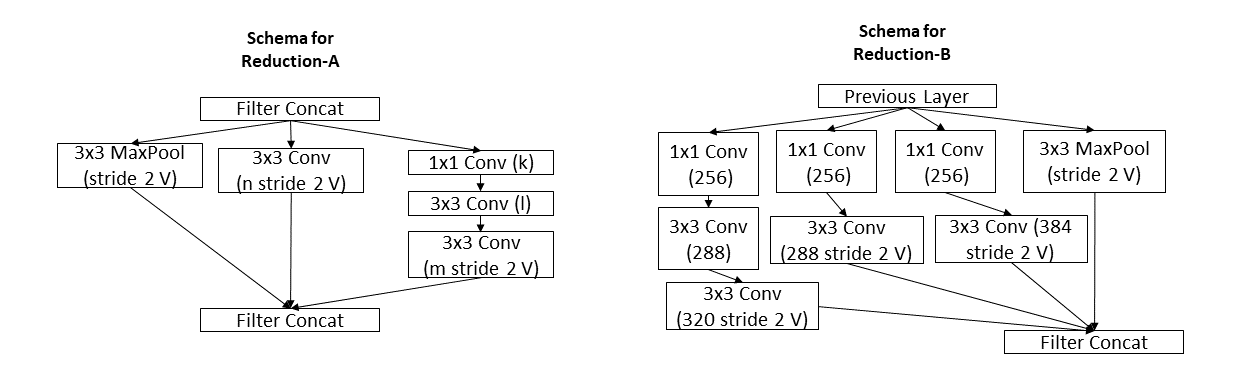}
        }%
    \end{center}
    \caption{%
        Inception-ResNet-V2 Architecture
     }%
   \label{fig:Inception-ResNet-V2-Architecture}
\end{figure}

Szegedy et al. \cite{szegedy2016inception} studied the combination of Inception architecture \cite{szegedy2016rethinking} and Residual connections \cite{he2016deep}, and proposed an architecture that is based on the Inception family of architectures by replacing the inception module with a hybrid Inception-ResNet module as shown in Figure~\ref{fig:Schema-for-Inception-ResNet-modules}, which are three variants: 1. Inception-ResNet-A for 35 x 35 grid, 2. Inception-ResNet-B for 17 x 17 grid, and 3. Inception-ResNet-C for 8 x 8 grid. Szegedy et al. \cite{szegedy2016inception} argued that training with residual connections significantly accelerated the training of Inception networks. 
The large scale schema structure and the detailed structure of its components are shown in Figure~\ref{fig:Inception-ResNet-V2-Architecture}. The input image of size 299 x 299 x 3 under-goes a series of convolutions in the Stem module, as shown in Figure~\ref{fig:Schema-for-Stem-module}, followed by the hybrid Inception-ResNet modules. Each hybrid Inception-ResNet module is followed by a Reduction module, as shown in Figure~\ref{fig:Schema-for-Reduction-modules} to reduce the dimensions of the representation. 
Later, the final hybrid Inception-ResNet module’s output is fed to the average pooling layer, followed by a dropout layer to output the predictions. The design of such deep neural networks that increases the number of layers leads to instability during training. 
The network may die early, for example. Szegedy et al. \cite{szegedy2016inception} suggested scaling down the residuals before adding them to the previous activation layer to stabilize the training, and He et al. \cite{he2016deep} suggested a two-phase training where the first warm-up phase is performed with a low learning rate and followed by a high learning rate in the second phase. Szegedy et al. \cite{szegedy2016inception} also trained an ensemble of one Inception-v4 and three Inception-ResNet-v2 models on the ILSVRC 2012 classification task (ImageNet dataset \cite{russakovsky2015imagenet}) and achieved 3.08\% top-5 error rate on the test set of the ImageNet dataset.

\subsubsection{ResNet152V2} \label{ResNet152V2}

\begin{figure}[htbp]
     \begin{center}
        \subfigure[ResNet152-V2 Architecture]{%
            \label{fig:ResNet152-V2-Architecture}
            \includegraphics[scale=0.5]{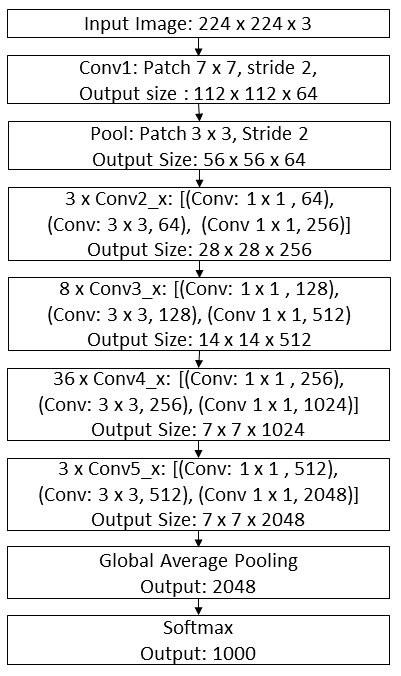}
        }%
        \quad\quad\quad\quad\quad\quad\quad\quad\quad\quad
        \subfigure[Residual Unit]{%
           \label{fig:Residual-Unit}
           \includegraphics[scale=0.5]{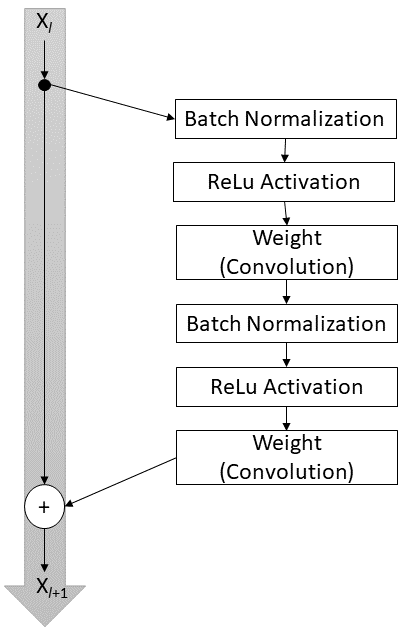}
        }\\ 
    \end{center}
    \caption{%
        ResNet152-v2 Architecture and its Residual Unit
     }%
   \label{fig:ResNet152-v2 Architecture and its Residual Unit}
\end{figure}

The Deep Residual Networks introduced by He et al. \cite{he2016deep} have improved the accuracy of the deep architecture models and are shown to have excellent convergence behaviors. 
He et al. \cite{he2016identity} studied the propagation formulation behind the residual blocks, i.e., to create a direct path for propagating information through the entire network including the residual unit as shown in Figure~\ref{fig:Residual-Unit} and demonstrated that when the identity maps are used as the skip connections and after-addition activation, forward and backward signals are directly propagated between any residual blocks. 
Identity mappings help protect the network from vanishing gradient problem. 
The significant difference between ResNet-V1 and ResNet-V2 is that before the convolution, ResNet-V2 performs batch normalization and ReLU activation at the input; whereas, ResNet-V1 performs convolution, followed by batch normalization and ReLU activation. 
The architecture of ResNet152-v2 is shown in Figure~\ref{fig:ResNet152-v2 Architecture and its Residual Unit}, which takes an input image of size 224 x 224 x 3 that goes through an initial convolution with a kernel size of 7 x 7 followed by a Pooling operation with a kernel size of 3 x 3.

Later, the pooling layer's output is passed on to a series of Residual blocks, each containing three layers: 1 x 1 Convolution, 3 x 3 convolution, and a 1 x 1 convolution, which is then followed by an Average Pooling layer and a fully connected layer with softmax activation to output the class of Imagenet dataset. 
When trained on Imagenet dataset \cite{russakovsky2015imagenet} with this architecture, He et al. \cite{he2016identity} reported the top-1 error rate of 21.1\% and  top-5 error rate of 5.5\%.

\subsubsection{DenseNet-201} \label{DenseNet201}
Computer Vision and Pattern Recognition (CVPR) is an annual international conference regarded as one of the field's most important and influential conferences. Densely Connected Convolutional Networks (DenseNet) introduced by Huang et al. \cite{huang2017densely} won the best paper award at the CVPR 2017 conference \cite{CVPR20175:online},  which connects each layer of the network in a feed-forward manner to every other layer.
\begin{figure}[htbp]
     \begin{center}
        \subfigure[DenseNet-201 Large Scale Architecture]{%
            \label{fig:DenseNet-201-Large-Scale-Architecture}
            \includegraphics[scale=0.5]{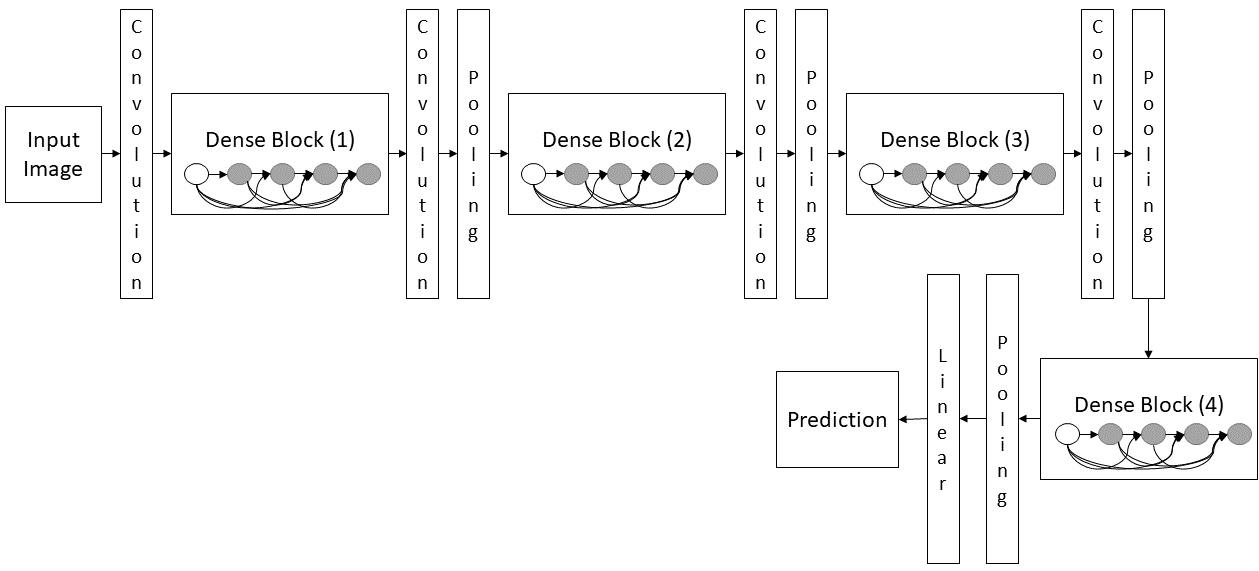}
        }\\
        \subfigure[DenseNet-201 Architecture]{%
           \label{fig:DenseNet201-Architecture}
           \includegraphics[scale=0.5]{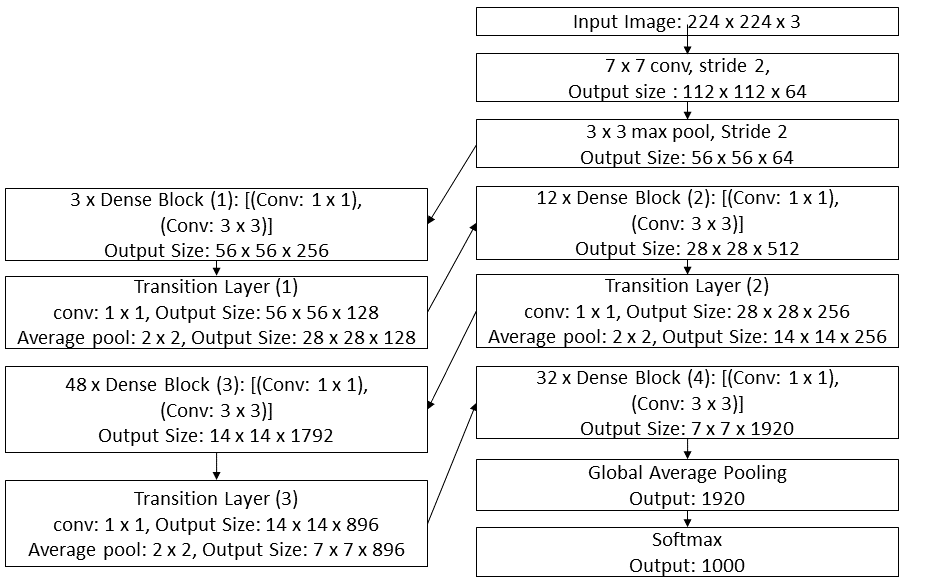}
        }%
        \subfigure[DenseNet-201 Convolution Block]{%
           \label{fig:DenseNet-201-Convolution-Block}
           \includegraphics[scale=0.5]{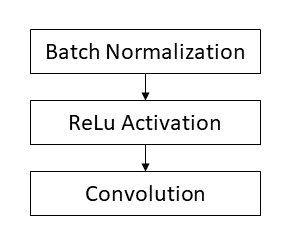}
          }%
    \end{center}
    \caption{%
        DenseNet-201 Architecture
     }%
   \label{fig:DenseNet-201-Architecture}
\end{figure}

DenseNets have a similar advantage to that of ResNets \cite{he2016deep}, \cite{he2016identity} in solving the problem of vanishing gradients and several other benefits, including enhancing the propagation of features between the layers, facilitating the re-use of features, and significantly reducing the overall learnable parameters of the network. 
The DenseNet-201 architecture is shown in Figure~\ref{fig:DenseNet-201-Architecture}, which takes an input image of size 224 x 224 x 3 that goes through an initial convolution of kernel size 7 x 7 and stride 2, followed by a Max pooling operation of kernel size 3 x 3 and stride 2. 
Later, the max-pooling output is subjected to a series of dense blocks and transition layers (four dense blocks and three transition layers). 
The dense block consists of a 1 x 1 convolution followed by a 3 x 3 convolution where each convolution operation is a sequence of Batch Normalization, ReLU Activation, and Convolution as shown in Figure~\ref{fig:DenseNet-201-Convolution-Block}. 
The transition layers have a sequence of 1 x 1 convolution followed by an average pooling of 2 x 2. At the end of the fourth dense block, the global average pooling is carried out with softmax activation. 
Huang et al. \cite{huang2017densely} reported that with only 0.8 million parameters (about 1/3 of ResNet parameters), the DenseNet-201 architecture is able to achieve a comparable accuracy of ResNet \cite{he2016identity} with 10.2 million parameters. 
When trained on the ImageNet dataset \cite{russakovsky2015imagenet}, the top-1 error rate was 22.58\%, and the top-5 error rate was 6.34\%.

\subsection{Classification Performance Metrics} \label{Classification Performance Metrics}
Evaluation metrics are critical for accessing the performance of a deep learning classification model.
There are different metrics of assessment that are available for these purposes. However, the standard metrics reported in the literature for deep learning classification tasks are accuracy, precision, recall, f1 score; all of them are calculated using the confusion matrix, as shown in the Figure~\ref{fig:confusion_matrix}, and area under the ROC (receiver operating characteristics) curve (AUC).

\begin{figure}[htbp]
    \begin{center}
    {

        \offinterlineskip

        \raisebox{-6cm}[0pt][0pt]{
            \parbox[c][5pt][c]{1cm}{\hspace{-4.1cm}\rot{\textbf{Actual}}\\[20pt]}}\par

        \hspace*{1cm}\MyHBox[\dimexpr3.4cm+6\fboxsep\relax]{Predicted}\par

        \hspace*{1cm}\MyHBox{Negative}\MyHBox{Positive}\par

        \MyTBox{\rot{Negative}}{True Negative (TN)}{False Positive (FP)}

        \MyTBox{\rot{Positive}}{False Negative (FN)}{True Positive (TP)}

    }
\end{center}
\caption{%
        Confusion Matrix
     }%
   \label{fig:confusion_matrix}

\end{figure}
\begin{center}
True Positive (TP): Predicted values correctly predicted as an actual positive\\
False Positive (FP): Predicted values incorrectly predicted an actual positive\\
False Negative (FN): Positive values predicted as negative\\
True Negative (TN): Predicted values correctly predicted as an actual negative\\
\end{center}

\paragraph{Accuracy} The accuracy of the model is calculated using the equation~\ref{accuracy}, which is a ratio of correct predictions to the total predictions.
\begin{equation}
\label{accuracy}
Accuracy=\frac {TP+TN}{TP+TN+FP+FN}
\end{equation}

\paragraph{Precision} The precision of the model summarizes model's accuracy in terms of the number of samples which were predicted positive and is given by the equation~\ref{precision}.
\begin{equation}
\label{precision}
Precision=\frac {TP}{TP+FP}
\end{equation}

\paragraph{Recall} Recall of the model is calculated using the equation~\ref{recall}, that tells how well the positive class was predicted.
\begin{equation}
\label{recall}
Recall=\frac {TP}{TP+FN}
\end{equation}

\paragraph{F1 Score} F1 score is the calculation of harmonic mean of precision and recall of the model and is given by the equation~\ref{f1score}

\begin{equation} 
\label{f1score}
    \textit{F1 Score}=2*\frac{Precision*Recall}{(Precision+Recall)}
\end{equation}

\paragraph{AUC score} AUC score is the measure of area covered by the receiver operating characteristics (ROC) curve. For a perfect classifier, the AUC score is 1.0

\subsection{Weighted-Average Ensemble} \label{Weighted-Average Ensemble}
\begin{figure}[htbp]
    \centering
    \includegraphics[scale=0.5]{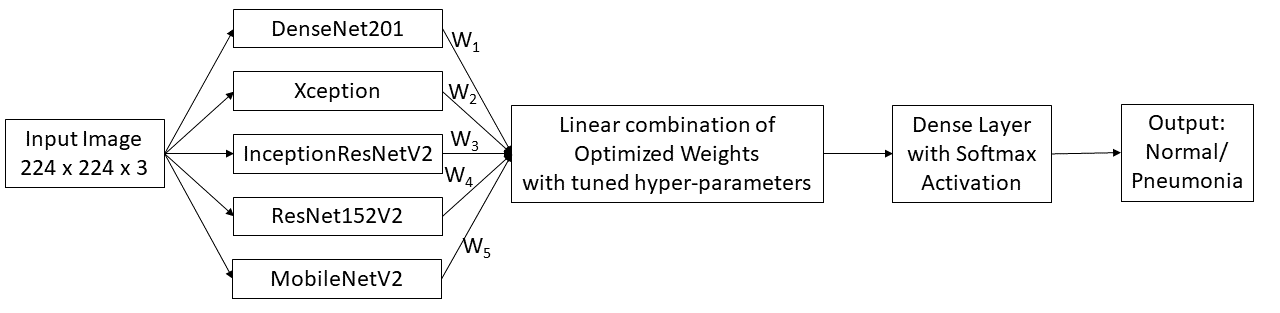}
    \caption{Weighted Average Ensemble}
    \label{fig:weighted_average_ensemble}
\end{figure}

Classification Algorithms based on a single architecture/ model often does not capture entire features in the data for optimal predictions.
The aggregation of multiple algorithms into an ensemble of models captures the data's underlying distribution more precisely, making better predictions \cite{shahhosseini2019optimizing}, \cite{brown2005diversity}, \cite{dietterich2000ensemble}.
The Figure~\ref{fig:weighted_average_ensemble} shows the building blocks of the weighted average ensemble model.

Each transfer learning model's output is then multiplied by a weight and then combined linearly, followed by a softmax layer to output predictions. 
During the training process, the weights are optimized with the condition that they add up to 1. These optimized weights determine the contribution of each transfer learning model in the final prediction.

\subsection{Dataset Description} \label{Dataset Description}
For all the experiments conducted in this study, we used a Chest X-ray dataset by Kermany et al. \cite{kermany2018identifying}. 
The dataset comprises 5,856 chest X-ray images taken from children are labeled either Normal or Pneumonia. 
The original training and test sets are heavily imbalanced. So, we initially combined the dataset with all Normal images in one folder and all the Pneumonia images in another folder. 

\begin{figure}[htbp]
     \begin{center}
        \subfigure[Normal Image]{%
            \label{fig:Normal_Image}
            \includegraphics[width=0.3\textwidth]{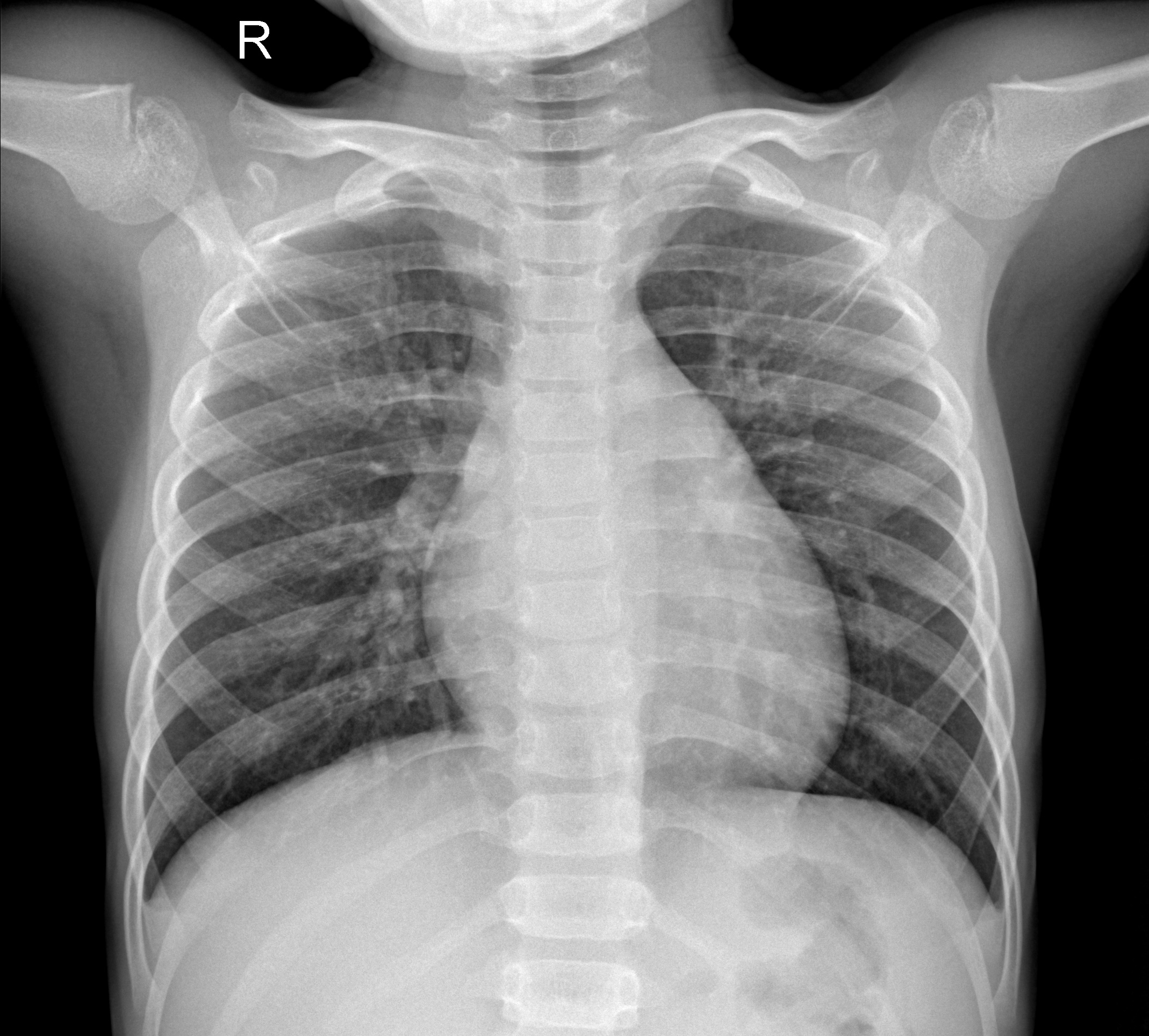}
        }
        \quad\quad\quad\quad\quad\quad\quad\quad\quad\quad
        \subfigure[Pneumonia Image]{%
           \label{fig:Pneumonia_Image}
           \includegraphics[width=0.3\textwidth]{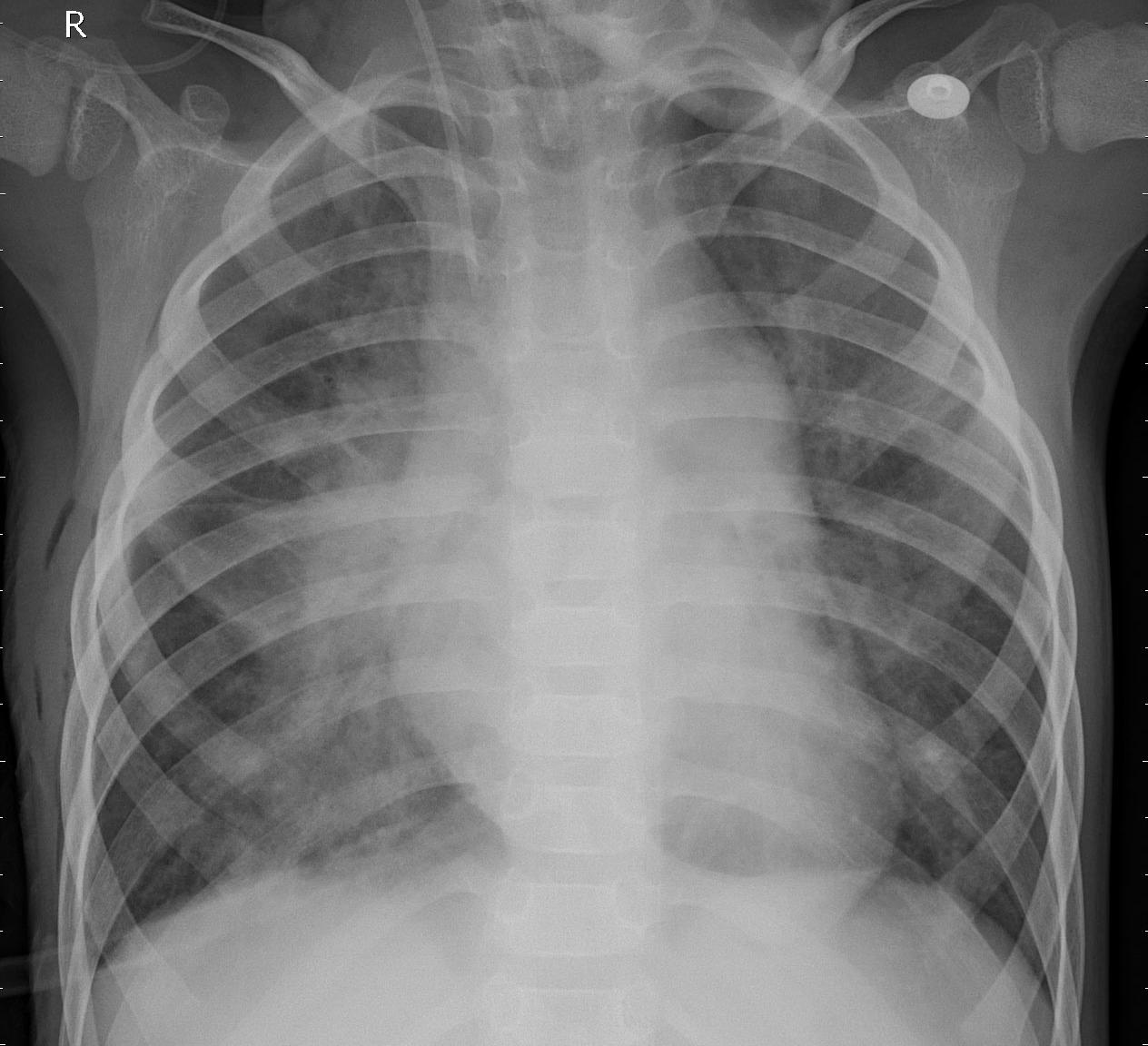}
        }
    \end{center}
    \caption{%
        Sample of a Normal and Pneumonia Image
     }%
   \label{fig:sampleimages}
\end{figure}

We show a sample of a Normal image and Pneumonia image in Figure~\ref{fig:sampleimages}. The dataset was then shuffled and split into training, validation, and test sets, of which 3,748 images in the training set, 936 images in the validation set, and 1,172 images in the test set.

\subsection{Data Preprocessing} \label{Data Preprocessing}
The chest X-ray images in the dataset are in varying sizes, i.e., all the chest X-ray images' dimensions are not the same. 
However, the deep neural network architectures utilized in this study as part of transfer learning expect all the images to be in a common dimension. 
For example, Xception architecture expects the dimensions of the image (width x height x no. of Channels) to be 299 x 299 x 3, and width and height should be no smaller than 71. 
The dimension of the input image will also vary by the type of deep neural network architecture. 
For example, the DenseNet201 architecture expects the input image shape to be (224 x 224 x 3), with width and height no smaller than 32, and InceptionResNet-V2 expects the input image shape to be (299 x 299 x 3), with width and height no smaller than 75. 
To have common dimensions accepted by all the architectures used in this study, we initially resized all the chest X-ray images to have the shape of (224 x 224 x 3).

\begin{figure}[htbp]
    \centering
    \includegraphics[width=0.3\textwidth]{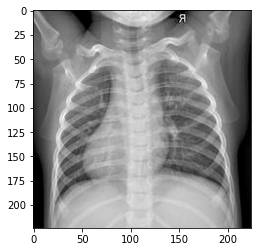}
    \caption{Chest X-ray image flipped left to right}
    \label{fig:random_flip}
\end{figure}

Once the images are resized to 224 x 224 x 3, we created TFRecords of the images and one-hot encoded the labels. 
TFRecord is a binary file format that is a standard and the most recommended data storage format in Tensorflow \cite{UseTPUsT54:online}. 
Storing data in a binary file format improves the data importing pipeline's performance and reduces the model's training time.
Deep learning models are in general data-hungry. 
They require a massive amount of data during training to capture the most relevant features; otherwise, the model does not generalize well when tested on new data. 
Data Augmentation is a technique used when the training data is limited to increase the training data size. 
This study augmented the training data by randomly flipping each image in a batch (see Figure~\ref{fig:random_flip}).

\subsection{Hyper-parameter tuning} \label{Hyper-parameter tuning}
Hyper-parameter tuning is one of the main contributions of this study. 
In the following subsections, we briefly discuss the parameters that are fine-tuned during training the model.
\subsubsection{Learning Rate} \label{Learning Rate}
The learning rate is one of the single most crucial hyper-parameters to be carefully chosen while training the model. 
In other words, it would be equally important to choose the appropriate learning rate for the model to select the right model from a family of models or learning algorithms.
The typical values of the learning rate while training a model with standardized inputs, i.e., the inputs are in the interval (0, 1), are greater than 1e-06 and less than 1 \cite{bengio2012practical}.

Since we are using transfer learning with pre-trained weights in this study, it is critical to have a very low learning rate to avoid the risk of overfitting very quickly.
High learning rates apply larger weight updates to the model. 
Therefore, it's best to avoid high learning rates as pre-trained models already hold decent weights that do not need larger weight updates again while using them as transfer learning models to train new datasets.
Other common strategies include learning rate warm-ups \cite{he2016deep}, \cite{goyal2017accurate} and reducing the learning rate on the plateau, which is a part of callbacks API in Keras \cite{ReduceLR67:online}.  
Learning rate warm-ups use less aggressive learning rates at the start of training. 
The other reduces the learning rate after a certain number of epochs if the model does not improve the monitored metrics, such as loss, accuracy, etc. during training.

In this study, the model's training started with a learning rate of 0.001 and then reduced the learning rate by a factor of 0.3 for every five epochs if the model did not improve.
This strategy worked better than others for the model to converge, where the last reported learning rate was 2.7e-05, which helped achieve the best model performance metrics.

\subsubsection{Batch Size} \label{batch_size}
Batch size is a configurable hyper-parameter during the training of a neural network model, which refers to the number of training examples used in a single iteration. 
Generally, the batch size is between 10 and 1000, and 32 is a good default value according to Bengio  \cite{bengio2012practical}.
The Tensor Processing Unit (TPU) was used to train the model, which consists of four processors, and each of them has two TPU cores, allowing eight cores for each TPU.
We set a batch size of 16 for each core of a TPU, with eight cores; the final batch size was 128.

\subsubsection{Number of Epochs} \label{epochs}
\begin{figure}[htbp]
    \centering
    \includegraphics[width=0.5\textwidth]{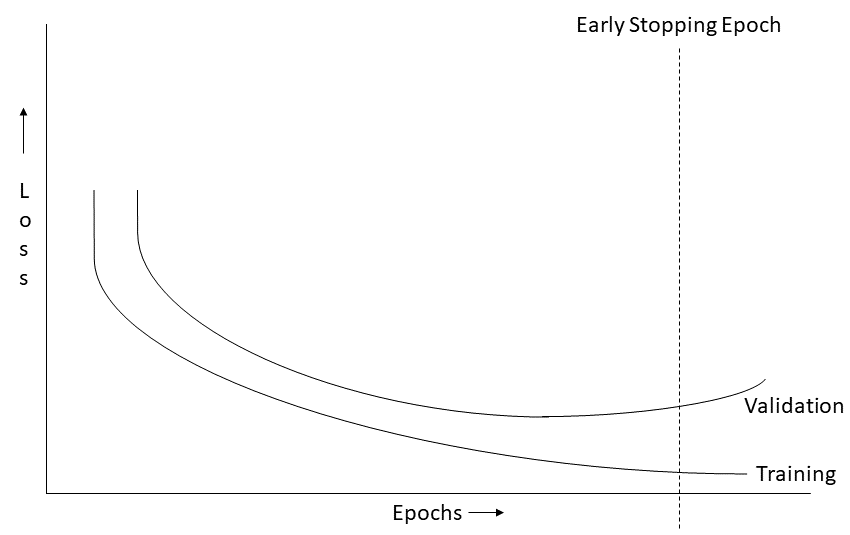}
    \caption{Early Stopping}
    \label{fig:early_stopping}
\end{figure}

The number of epochs or the number of training iterations is another hyper-parameter that can be optimized using the principle of Early stopping \cite{bengio2012practical}. 
Early stopping is another way of ensuring that the model does not overfit the training data by stopping the training process (see Figure~\ref{fig:early_stopping}), even though other hyper-parameters such as learning rate and batch size would yield over-fitting.
Early stopping comes as a callbacks API in Keras \cite{EarlySto4:online}. 
The patience parameter and the quantity to be monitored are set to 20 and the loss. 
When the model shows no improvement in the loss for 20 consecutive epochs, the compiler terminates the training process.

\subsection{Loss Function} \label{loss_function}
The dataset has two classes (Normal, Pneumonia), which is generally considered a binary classification problem(Normal - 0, Pneumonia - 1). 
In a binary classification task, we optimize the binary cross-entropy function, so the model spits out whether the chest X-ray image is Normal (0) or Pneumonia (1). 
In this study, we modeled the algorithm to spit out the probability of the image being Normal or Pneumonia.
So, we intend to optimize the categorical cross-entropy function instead of the binary cross-entropy function.

The categorical cross-entropy loss, also known as log loss or logistic loss or softmax loss, is given by the equation~\ref{cce}, where $M$ is the number of training examples, $K$ is the number of classes, ${y_{m}^{k}}$ is the target label for training example $m$ for class $k$, $x$ is the input for training example $m$, and $h_{\theta}$ is the model with neural network weights $\theta$. 

\begin{equation} \label{cce}
J_{cce}=-\frac {1}{M}\sum \limits _{k=1}^{K} \sum \limits _{m=1}^{M} {y_{m}^{k}\times \log \left ({h_{\theta }\left ({x_{m},k }\right) }\right)}
\end{equation}

The predicted class probabilities are compared with the actual classes/ labels (Normal, Pneumonia) to minimize the loss. The loss is calculated that penalizes for any deviation between the actual class and the model's output.
The penalty is a logarithmic loss that yields larger scores for larger deviations, which tends to 1, and smaller scores for small deviations tend to 0. A perfect model will have a categorical cross-entropy loss of 0.

\subsection{Optimization Algorithm} \label{optimization_algorithm}
After all the data preprocessing and hyper-parameters configuration, the next challenging task is choosing the right optimization algorithm from a pool of optimization algorithms, consisting of Gradient Descent (GD), Stochastic Gradient Descent (SGD), Adam, etc. 
Gradient Descent is the oldest and the traditional optimization algorithm that solves the optimal value along the gradient descent, converging at a linear rate.
In this method, the gradients of all the samples are calculated for each parameter update making the gradient descent cost calculation very high \cite{sun2019survey}.
To overcome this issue, Robbins and Monro \cite{robbins1951stochastic} proposed the Stochastic Gradient Descent (SGD) optimization method. 
In this method, the parameter updates are calculated using a random sample from a mini-batch that converge at a sub-linear rate. Even though the cost calculation is improved, choosing an appropriate learning rate is often challenging.
Kingma and Ba \cite{kingma2014adam} introduced Adam (Adaptive Moment Estimation), a stochastic optimization algorithm based only on first-order gradients. 
The algorithm improves the cost calculation with little memory and calculates individual adaptive learning rates for different parameters from the estimates of gradients' first and second moments.
The gradient descent process of the Adam optimization method is relatively stable compared to gradient descent and stochastic gradient descent methods and is most suitable for large datasets or parameters \cite{kingma2014adam}.
So, we used Adam as an optimization algorithm in this study.


\section{Results} \label{Results}

\subsection{Classification Performance Metrics} \label{Classification Performance Metrics Results}

After finalizing the hyper-parameter configurations and optimization algorithm, the models are compiled and fine-tuned during the training. 
The models' performance is evaluated on the test dataset, which consists of 1,172 chest X-ray images, and the confusion matrix is computed for each transfer learning model consisting of True Negatives, False Positives, False Negatives, and True Positives as shown in Table~\ref{tab:confusion_metrics}.
The Xception architecture performance is better than all other transfer learning architectures, while the weighted average ensemble outperformed every transfer learning model, including the Xception architecture.

\begin{table}[htbp]
    \centering
        \begin{tabularx}{\textwidth} {
          |>{\centering\arraybackslash}X
           >{\centering\arraybackslash}X
           >{\centering\arraybackslash}X
           >{\centering\arraybackslash}X
           >{\centering\arraybackslash}X| }
        \hline
         \textbf{Model} & \textbf{True Negative (TN)} & \textbf{False Positive (FP)} & \textbf{False Negative (FN)} & \textbf{True Positive (TP)} \\
        \hline
        \textbf{DenseNet201} & 303 & 14 & 7 & 848\\
        \textbf{Xception} &  302 & 15 & 5 & 850\\
        \textbf{InceptionResNet} & 303 & 14 & 7 & 848\\
        \textbf{ResNet152V2} & 299 & 18 & 9 & 846\\
        \textbf{MobileNetV2} & 303 & 14 & 7 & 848\\
        \hline
        \textbf{Ensemble Model} & \textbf{303} & \textbf{14} & \textbf{4} & \textbf{851}\\
        \hline
        \end{tabularx}
    \caption{Confusion Metrics}
    \label{tab:confusion_metrics}
\end{table}

As mentioned in Section~\ref{Classification Performance Metrics}, the accuracy, precision, recall, and f1 score are calculated for each transfer learning model (see Table~\ref{tab:classification_performance_metrics}). 
It is worth noting that the results of MobileNetV2 architecture are comparable to the best-performing architecture, i.e., the Xception architecture with approximately 20 million trainable parameters, which is almost ten times the MobileNet architecture. 
However, with about 162 million trainable parameters, the weighted average ensemble model outperformed all other models with test loss of 0.08 and achieving an accuracy of 98.46\%, precision of 98.38\%, recall of 99.53\%, f1 score of 98.96\%, and AUC of 99.60\%.

\begin{table}[htbp]
\centering
\begin{tabular}{|m{2.5cm} c c c c c c p{2cm}|}
\hline
 \textbf{Model} & \textbf{Accuracy} & \textbf{Precision} & \textbf{Recall} & \textbf{F1 Score} & \textbf{AUC} & \textbf{Test Loss} & \textbf{Total trainable parameters} \\
\hline
\textbf{DenseNet201} & 98.21 & \textbf{98.38} & 99.18 & 98.78 & 99.40 & 0.09 & 18,096,770 \\
\textbf{Xception} & 98.30 & 98.27 & 99.42 & 98.84 & 99.42 & 0.11 & 20,811,050\\
\textbf{InceptionResNet} & 98.21 & \textbf{98.38} & 99.18 & 98.78 & 99.38 & 0.09 & 54,279,266\\
\textbf{ResNet152V2} & 97.7 & 97.92 & 98.95 & 98.43 & 99.33 & 0.11 & 58,192,002\\
\textbf{MobileNetV2} & 98.21 & \textbf{98.38} & 99.18 & 98.78 & 99.08 & 0.11 & \textbf{2,226,434}\\
\hline
\textbf{Ensemble Model} & \textbf{98.46} & \textbf{98.38} & \textbf{99.53} & \textbf{98.96} & \textbf{99.60} & \textbf{0.08} & 162,638,991\\
\hline
\end{tabular}
\caption{Classification Performance Metrics}
\label{tab:classification_performance_metrics}
\end{table}

As mentioned in Section~\ref{Weighted-Average Ensemble}, the weights are optimized during training and the individual model weights are shown in Table\ref{tab:weighted_average_ensemble_model_weights}. 

\begin{table}[htbp]
\centering
\begin{tabular}{|c c|}
        \hline
         \textbf{Model} & \textbf{Weights}\\
        \hline
        \textbf{DenseNet201} & 0.22 \\
        \textbf{Xception} & 0.29\\
        \textbf{InceptionResNet} & 0.18\\
        \textbf{ResNet152V2} & 0.17\\
        \textbf{MobileNetV2} & 0.15\\
        \hline
        \end{tabular}
    \caption{Weighted Average Ensemble Model Weights}
    \label{tab:weighted_average_ensemble_model_weights}
\end{table}

The Xception and DenseNet201 architectures account for more than 50\% of the final predictions, with Xception architecture contributing 29\% of the final prediction and DenseNet201 architecture contributing 22\% of the final prediction.

\subsection{Comparison of results with other recent similar works} \label{Comparison of results with other recent similar works}
In this section, we compare the results from our study with other recent similar works (see Table~\ref{tab:comparision_of_results_with_othre_reent_similar_works}, best performance metrics are in bold). 
The results of our weighted average ensemble model outperformed all the classification metrics such as accuracy, precision, and f1 score, but recall and AUC from the comparable works to accurate classification of pneumonia.

\begin{table}[htbp]
\centering
\begin{tabular}{|c c c c c c|}
\hline
 \textbf{} & \textbf{Accuracy} & \textbf{Precision} & \textbf{Recall} & \textbf{F1 Score} & \textbf{AUC}\\
\hline
\textbf{Kermany et al. \cite{kermany2018identifying}} & 92.80 & 87.20 & 93.20 & 90.10 & 96.80\\
\textbf{Nahid et al. \cite{nahid2020novel}} & 97.92 & \textbf{98.38} & 97.47 & 97.97 & - \\
\textbf{Stephen et al. \cite{stephen2019efficient}} & 93.73 & - & - & - & - \\
\textbf{Chouhan et al. \cite{chouhan2020novel}} & 96.39 & 93.28 & \textbf{99.62} & 96.35 & 99.34 \\
\textbf{Rajaraman et al. \cite{rajaraman2018visualization}} & 96.20 & 97.00 & 99.50 & - & 99.00 \\
\textbf{Hashmi et al. \cite{hashmi2020efficient}} & 98.43 & 98.26 & 99.00 & 98.63 & \textbf{99.76} \\
\textbf{Mittal et al. \cite{mittal2020detecting}} & 96.36 & - & - & - & - \\
\textbf{Rahman et al. \cite{rahman2020transfer}} & 98.00 & 97.00 & 99.00 & 98.10 & 98.00 \\
\hline
\textbf{Current Work} & \textbf{98.46} & \textbf{98.38} & 99.53 & \textbf{98.96} & 99.60\\

\hline
\end{tabular}
\caption{Comparison of results with other recent similar works}
\label{tab:comparision_of_results_with_othre_reent_similar_works}
\end{table}


\section{Conclusions} \label{Conclusions}
According to the World Health Organization (WHO), pneumonia is one of the world's largest infectious cause of death in children, particularly children under the age of five \cite{Pneumoni94:online} and Centers for Disease Control and Prevention (CDC) estimates that pneumonia is one of the leading causes of death among adults in the United States \cite{Pneumoni23:online}. 
Chest X-rays are the standard technique used by radiologists in detecting pneumonia, and even for the well-trained radiologist, it is not uncommon to overlook pneumonia detection.
Due to the challenges of obtaining massive training data mainly because of high annotation costs, we used transfer learning techniques combined with data augmentation to overcome overfitting during the model training process.
This study proposes a weighted average ensemble model by fine-tuning the deep transfer learning architectures to improve the classification performance metrics such as accuracy, precision, recall, and f1 score to detect pneumonia from chest X-ray images.
To the best of our knowledge, we achieved the best classification performance metrics ever reported in the literature for pneumonia classification with accuracy of 98.46\%, precision of 98.38\%, and f1 score of 98.96\%.

\subsection*{Acknowledgements}

We thank Harrisburg University of Science and Technology for their support.

\bibliographystyle{ieeetr}
\bibliography{references} 
\end{document}